\documentclass{aa}
\usepackage{graphicx}
\usepackage{txfonts}
\usepackage{natbib}
\bibpunct{(}{)}{;}{a}{}{,}
\begin{document}

\title{A two-zone model for the emission from \object{RX~J1713.7$-$3946}}

\author{K. Moraitis \and A. Mastichiadis}

\offprints{K. Moraitis, \email{kmorait@phys.uoa.gr}}

\institute{Department of Physics, University of Athens,
Panepistimiopolis, 15783 Zografos, Athens }

\date{Received ... / Accepted ...}

\abstract{} {We study the acceleration and radiation of charged
particles in the shock waves of supernova remnants using a recent
version of the "box model". According to this, particles are
accelerated in an energy--dependent region around the shock by the
first order Fermi mechanism and lose energy through radiation.} {The
particle distribution function is obtained from a spatially averaged
kinetic equation that treats the energy losses self--consistently.
There exists also a second population that consists of those
particles that escape behind the shock where they also radiate. The
energy distribution of this population is calculated in a similar
manner.} {The application of the model to the supernova remnant
\object{RX~J1713.7$-$3946}, which was recently confirmed as a TeV
source by H.E.S.S., shows that the X-ray emission can be attributed
to electron synchrotron radiation while in $\gamma$-rays there are
contributions from both electrons and protons, with protons playing
the dominant role. Additionally, there are strong indications that
particles diffuse in turbulence that has a Kolmogorov spectrum.}{}


\maketitle

\section{Introduction}\label{sec:intro}

Supernova remnants (SNRs) have long thought to be the sites of
galactic cosmic ray acceleration, at least up to energies close to
that of the "knee" ($\sim10^{15}\ \mathrm{eV}$) of their spectrum
\citep{krymskii77,bell78,blandford78}. Direct evidence for this came
only during the last decade with the observation of non-thermal
X-ray emission from SNRs. This can be best explained as synchrotron
radiation by electrons that have been accelerated up to energies
$\sim100\ \mathrm{TeV}$ \citep{reynolds99}. Despite this success,
evidence for the presence of a relativistic hadronic component is
less certain. Accelerated hadrons in SNRs are expected to emit high
energy gamma rays from the decay of neutral pions ($\mathrm{\pi}^0$)
produced in proton-proton collisions \citep{dorfi91}. These are best
observed in the TeV regime \citep{drury94}, thus SNRs have been one
of the main targets for the ground based \v{C}erenkov detectors
\citep[for a review see][]{weekes02}. However, the above picture is
complicated from the fact that the same high energy electrons which
produce the synchrotron X-rays will Compton up-scatter the microwave
photons to TeV energies \citep{mastichiadis96,pohl96}, contributing
an extra component in this regime. Therefore one expects that any
TeV detection of a SNR cannot be unequivocally attributed to the
presence of a relativistic hadronic (or leptonic) component. Instead
careful modeling is required to determine the origin of these
gamma-rays.

In modeling the multi-wavelength spectrum of SNRs most authors
assume for the particle energy distribution (electrons or protons)
the form of a power-law with an exponential cut-off. The power-law
is a natural consequence of Fermi acceleration while the exponential
turnover is supposed to arise from energy losses or SNR age
limitations. Despite the theoretical support for these assumptions,
this approach is still phenomenological. On the other hand, the use
of extensive numerical codes has proved a valuable tool for studying
the acceleration, especially as they allow more complex phenomena,
such as shock modification, to be taken into account
\citep{ellison01,berezhko02}. The application of an analytical model
however is interesting as it offers a good opportunity to get an
insight at some of the basic principles of particle acceleration.

A development in the direction of an analytical approach to the
problem of particle acceleration has been made recently by
\citet{drury99}. This approach has expanded on the concept of the
so-called "box" model for particle acceleration \citep[for a review
see][]{kirk94} by assuming an energy-dependent size for the
acceleration region. This inserts some modifications on the particle
spectra and is thus interesting to apply this model to fit the SNR
multi-wavelength emission. A good starting point for this is the
first SNR with confirmed TeV emission, \object{RX~J1713.7$-$3946}
\citep{muraishi00,enomoto02,aharonian04,aharonian05}.

\object{RX~J1713.7$-$3946} (\object{G347.3$-$0.5}) is a shell-type
supernova remnant that was discovered in X-rays during the ROSAT All
Sky Survey \citep{pfeffermann96}. It is a young SNR with an age of
$\sim2000$ years, possibly associated with the supernova that took
place in AD393 \citep{wang97} and is located in the Galactic plane
at a distance $\sim1\ \mathrm{kpc}$. The X-ray emission of
\object{RX~J1713.7$-$3946} is purely non-thermal in nature
\citep{koyama97,slane99} and its X-ray morphology, as revealed by
Chandra, is quite complicated consisting of bright filaments and
dark voids \citep{uchiyama03} indicating spatially inhomogeneous
magnetic field and/or density of relativistic particles. In $\gamma
$-rays, the remnant was originally observed by the CANGAROO
telescope \citep{muraishi00,enomoto02} and recently confirmed as a
high energy source by H.E.S.S. \citep{aharonian04,aharonian05}. The
remnant also exhibits a weak radio emission \citep{lazendic04}. The
broadband energy spectrum of \object{RX~J1713.7$-$3946} has been
modeled by many authors, assuming either a leptonic
\citep{uchiyama03,pannuti03,lazendic04} and/or a hadronic origin
\citep{enomoto02,aharonian05,berezhko06} for the $\gamma$-ray
emission, but without a conclusive result so far.

In this paper we apply the \citet{drury99} model to the supernova
remnant \object{RX~J1713.7$-$3946} and draw some conclusions about
its properties. In section~\ref{sec:model} we review the main
characteristics of the model and obtain the distribution function of
accelerated particles as well as that of the particles which have
escaped behind the shock. In section~\ref{sec:apply} we derive the
expected multi-wavelength photon spectrum from these distributions
and apply it to \object{RX~J1713.7$-$3946}. Finally, in
section~\ref{sec:discuss} we summarize and discuss our results.

\section{The acceleration model}\label{sec:model}

Charged particles interacting with the shock wave of a supernova
remnant can be accelerated to high energies through first order
Fermi acceleration. Alfv\'{e}n waves in the background medium
scatter the charged particles, changing their direction but not
their energy. Particles are thus confined to a region around the
shock and are forced to cross it many times. The difference in the
gas velocity upstream and downstream results in the statistical
acceleration of particles after many shock crossings. While confined
around the shock particles also lose energy because they emit
radiation. At the same time a fraction of them escapes behind the
shock towards the interior of the remnant where it also radiates. We
next calculate the distribution function of these two populations
when the accelerated particles are electrons as well as when they
are protons. We denote quantities in the acceleration region with
the subscript 1 and in the escape region with the subscript 2.

Although the shock in a SNR is spherical we can treat the
acceleration in the plane shock geometry as long as the size of the
acceleration region is much smaller than the SNR radius. We assume
the presence of a uniform magnetic field that is normal to the
shock. We also adopt the test particle approximation, i.e. we assume
zero particle pressure. With these assumptions, acceleration takes
place in a region around the shock of size
$L=L_{\mathrm{u}}+L_{\mathrm{d}}=K_{\mathrm{u}}/U_{\mathrm{u}}+K_{\mathrm{d}}/U_{\mathrm{d}}$
where $K_{\mathrm{u(d)}}$, $U_{\mathrm{u(d)}}$ are the upstream
(downstream) diffusion coefficient and gas velocity respectively.
Assuming that the diffusion coefficient is continuous across the
shock, i.e. $K_{\mathrm{u}}=K_{\mathrm{d}}=K$, the size of the
acceleration region can be further written $L=K(r+1)/U_{\mathrm{u}}$
where $r=U_{\mathrm{u}}/U_{\mathrm{d}}$ is the compression ratio of
the shock. Following \citet{drury99} we let the diffusion
coefficient to be momentum-dependent and adopt a general power-law
dependence $K(p)=\kappa p^\delta $ with $\delta >0$. This choice
makes the mathematical analysis easier but also covers some
physically interesting cases. For example, the case $\delta=1$
corresponds to Bohm diffusion, $\delta=1/3$ to diffusion from
Alfv\'{e}n waves with a Kolmogorov spectrum of turbulence and
$\delta=1/2$ to a Kraichnan spectrum. The same power-law dependence
holds for the size of the acceleration region $L(p)=\kappa
(r+1)p^\delta /U_{\mathrm{u}}$ as well as the acceleration timescale
which is given by the relation
\begin{equation}\label{eq:tacc}
t_{\mathrm{acc}}(p)=\frac{3L(p)}{U_{\mathrm{u}}-U_{\mathrm{d}}}.
\end{equation}

If the accelerated particles are electrons they also experience
energy losses through synchrotron radiation and/or inverse Compton
scattering (IC). In the case where the latter is in the Thomson
limit, we can write for the particles total losses
$\dot{p}=-\alpha_{1} p^2$, where
\begin{equation}\label{eq:alpha}
\alpha_{1}
=\frac{4}{3}\frac{\sigma_{\mathrm{T}}(U_{\mathrm{B_{1}}}+U_{\mathrm{ph}})}{m^{2}c^{2}}
\end{equation}
is essentially the sum of the energy densities of the magnetic field
$U_{\mathrm{B_{1}}}$ and the target photon field $U_{\mathrm{ph}}$,
$\sigma_{\mathrm{T}}$ is the Thomson cross section, $m$ the electron
mass and $c$ the speed of light. If there were no energy losses,
particles would escape from the acceleration region with the
downstream flow at the bulk velocity $U_{\mathrm{d}}$. The presence
of losses however generates a downward flux in momentum space and
when combined with the increasing size of the acceleration region
gives an additional escape process. The velocity of escape thus
increases and the escape timescale is given by
\begin{equation}\label{eq:tesc}
t_{\mathrm{esc}}(p)=\frac{L(p)}{U_{\mathrm{d}} + \alpha_{1} p^2
\frac{dL_{\mathrm{d}}}{dp}}.
\end{equation}
The second term in the denominator is the addition of the
\citet{drury99} model to the standard box model and vanishes for
constant diffusion coefficient or no energy losses.

The evolution of the spatially averaged isotropic distribution
function of particles at the shock front, $N_{1}(p,t)$, is governed
by the continuity equation
\begin{equation}\label{eq:dndt}
\frac{\partial N_{1}}{\partial t}+\frac{\partial }{\partial p}\left[
N_{1}\left(\frac{p}{t_{\mathrm{acc}}(p)}-\alpha_{1}
p^2\right)\right]+\frac{N_{1}}{t_{\mathrm{esc}}(p)}=Q(p,t).
\end{equation}
The first term in brackets describes acceleration at the momentum
dependent rate $t_{\mathrm{acc}}^{-1}$ and the second the
synchrotron-type energy losses. The last term in the left hand side
describes particle escape at the momentum dependent rate
$t_{\mathrm{esc}}^{-1}$ and the source term
$Q(p,t)=Q_{\mathrm{o}}\delta (p-p_{\mathrm{o}})\vartheta (t)$
represents injection of particles at low momenta $p_{\mathrm{o}}$
with the constant rate $Q_{\mathrm{o}}$; $\delta (x)$ and $\vartheta
(x)$ are the Dirac delta and Heavyside (or unit-step) functions
respectively.

\begin{figure}
\resizebox{\hsize}{!}{\includegraphics{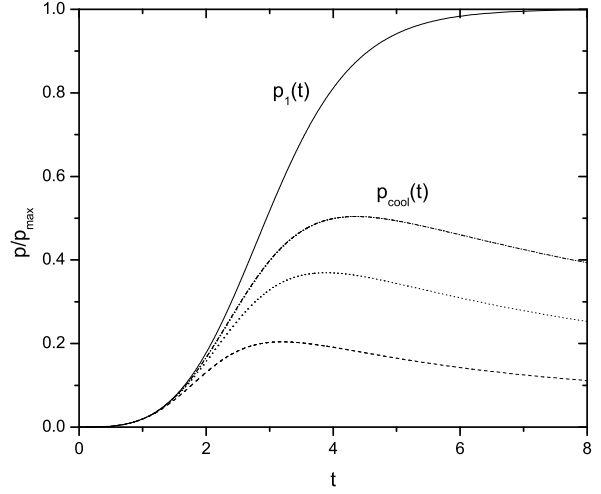}} \caption{The
evolution of the upper limit of the particle distribution function
$p_{1}(t)$ (solid line) for parameters $p_\mathrm{o}=0$,
$\delta=0.3$ and $B_1=10\ \mathrm{\mu G}$. Time is measured in units
of the cooling timescale $t_{\mathrm{cool,1}}=2\,200\
(p_{\mathrm{max}}\ c/50\ \mathrm{TeV})^{-1}\mathrm{yr}$. The
evolution of the cooling momentum $p_{\mathrm{cool}}(t)$ for the
cases $B_2=10\ \mathrm{\mu G}$ (dashed line), $B_2=5\ \mathrm{\mu
G}$ (dotted line) and $B_2=2\ \mathrm{\mu G}$ (dash-dotted line) is
also shown.} \label{char}
\end{figure}

From the equation describing the evolution of a particle's momentum
\begin{equation}\label{eq:char}
\frac{dp}{dt}=\frac{p}{t_{\mathrm{acc}}(p)}-\alpha_{1} p^2
\end{equation}
it follows that particles can be accelerated up to the momentum
where the acceleration and cooling timescales are equal, namely
\begin{equation}\label{eq:pstar}
p_{\mathrm{max}}=\left[\frac{(r-1)U_{\mathrm{u}}^2}{3r(r+1)\alpha_{1}
\kappa }\right] ^{\frac{1}{\delta +1}}.
\end{equation}
Particles reach this maximum allowed momentum if given enough time,
otherwise their momentum is determined from the solution of Eq.
(\ref{eq:char}). This equation can be solved analytically for the
time a particle of initial momentum $p_{\mathrm{o}}$ needs to reach
momentum $p$ and the solution is
\begin{equation}\label{eq:ptout}
t_1(p)=t_{\mathrm{cool,1}}\left[ \frac{x^{\delta }}{\delta }\
_2F_1\left(\frac{\delta }{\delta +1},1;1+\frac{\delta }{\delta
+1};x^{\delta +1}\right)
\right]^{p/p_{\mathrm{max}}}_{p_{\mathrm{o}}/p_{\mathrm{max}}},
\end{equation}
where $_2F_1$ is the hypergeometric function and
$t_{\mathrm{cool,1}}=(\alpha_{1} p_{\mathrm{max}})^{-1}$ is the
minimum cooling timescale. The inverse function of $t_1(p)$ gives
the maximum momentum of a particle at time $t$ and is denoted as
$p_1(t)$. This maximum momentum increases with time until energy
losses become important and the acceleration saturates at
$p_{\mathrm{max}}$ as can be seen in Fig.~\ref{char}. The time
required for this (in units of $t_{\mathrm{cool,1}}$) is a
decreasing function of the index $\delta$.

The solution of Eq. (\ref{eq:dndt}) for the distribution function is
then
\begin{eqnarray}\label{eq:npt}
N_1(p,t)&&=\frac{Q_{\mathrm{o}}t_{\mathrm{a,o}}}{p_{\mathrm{o}}}
\left(\frac{p}{p_{\mathrm{o}}}\right)^{-s_1}\nonumber\\
&&\left[1-\left(\frac{p}{p_{\mathrm{max}}}\right)^{\delta
+1}\right]^{-b}\left[1-\left(\frac{p_{\mathrm{o}}}{p_{\mathrm{max}}}\right)^{\delta+1}\right]^{b-1}
\end{eqnarray}
in the momentum range $p_{\mathrm{o}}<p<p_{1}(t)$ and zero
otherwise; $t_{\mathrm{a,o}}$ is the minimum acceleration timescale,
$t_{\mathrm{acc}}(p_{\mathrm{o}})$, and the index $b=\frac{1}{\delta
+1}\left( \frac{r-4}{r-1}+\frac{\delta }{r+1}\right)$. The
accelerated particles form a power-law of index
$s_1=\frac{r+2}{r-1}-\delta $, smaller by $\delta $ from the index
predicted by the standard box model, that extends from the injection
momentum $p_{\mathrm{o}}$ up to the maximum momentum $p_1(t)$. Given
enough time the particle distribution function reaches steady-state
and is then given by Eq. (\ref{eq:npt}) in the momentum range
$p_{\mathrm{o}}<p<p_{\mathrm{max}}$. In steady-state the power-law
gets modified since the first term in brackets in Eq. (\ref{eq:npt})
is then zero at $p=p_{\mathrm{max}}$ and a pile-up or a cut-off of
particles forms depending on whether the index $b$ is positive or
negative respectively.

Particles that escape towards the interior of the remnant play the
most important role in the emission \citep{ball92,aharonian99} and
we calculate next their energy distribution. The distribution
function of escaping particles obeys a similar equation to Eq.
(\ref{eq:dndt}) where now the source term is the escape rate of
accelerated particles while there are no acceleration or escape
terms
\begin{equation}\label{eq:dndt2}
\frac{\partial N_{2}}{\partial t}-\frac{\partial }{\partial p}\left(
\alpha_{2} p^2N_{2}\right)=\frac{N_1(p,t)}{t_{\mathrm{esc}}(p)}.
\end{equation}
The energy losses have the same form as in the acceleration zone
(see Eq. \ref{eq:alpha}) but depend on the magnetic field of the
escape zone, $B_2$. The solution of Eq. (\ref{eq:dndt2}) can be
easily obtained numerically (and semi-analytically) but we prefer a
simplified analytic approach. If the energy losses are negligible
the solution is simply the escape rate integrated over time
\begin{eqnarray}\label{eq:npt2}
N_2(p,t)&&=\frac{Q_{\mathrm{o}}\,\left(t-t_1(p)\right)}{p_{\mathrm{o}}}\left(
\frac{p}{p_{\mathrm{o}}}\right)^{-s_2}\left[
1-\left(\frac{p}{p_{\mathrm{max}}}\right)^{\delta
+1}\right]^{-b}\nonumber\\
&&\left[1-\left(\frac{p_{\mathrm{o}}}{p_{\mathrm{max}}}\right)^{\delta
+1}\right]^{b-1}\left[\frac{3}{r-1}+\delta \frac{r}{r+1}
\left(\frac{p}{p_{\mathrm{max}}}\right)^{\delta +1}\right],
\end{eqnarray}
which, for $p\ll p_1(t)$, is a power-law with the standard box model
index $s_2=s_1+\delta =\frac{r+2}{r-1}$. This result remains valid
when losses are present but only for those particles that had not
enough time to cool, i.e. those with momentum less than
\begin{equation}\label{eq:pcool}
p_{\mathrm{cool}}(t)=\frac{p_1(t)}{1+\alpha_{2}\,p_1(t)\,t}.
\end{equation}
This equation describes the evolution of momentum of the highest
energy particles as they cool in the escape zone and is shown in
Fig.~\ref{char}. On the other hand, particles with momentum
exceeding $p_{\mathrm{cool}}(t)$ had the necessary time to cool and
their distribution is steeper from the injected one by one power of
$p$, i.e.
\begin{equation}\label{eq:npt2cl}
N_{2,\mathrm{cool}}(p,t)\simeq
N_2(p,t)\frac{p_{\mathrm{cool}}(t)}{p}.
\end{equation}
The escaping particles thus have a broken power-law distribution,
given by Eq. (\ref{eq:npt2}) for $p<p_{\mathrm{cool}}(t)$ and by Eq.
(\ref{eq:npt2cl}) for $p_{\mathrm{cool}}(t)<p<p_{1}(t)$. The break
occurs at the cooling momentum which depends on the magnetic field
of the escape region. The extent of the cooled part of the particle
distribution, $p_1(t)-p_{\mathrm{cool}}(t)$, is greater for higher
values of $B_2$ as is shown in Fig.~\ref{char}. The steady-state
escape distribution consists only of cooled particles and hence has
a power-law index equal to $s_{2}+1$.

We next consider the case that the accelerated particles are
protons. Since protons lose substantial energy only through the
production of pions and the corresponding cooling timescale is much
larger than the age of the remnant, $t_{\mathrm{pp}}\simeq
5\,10^{15}\ \mathrm{s}\gg t_{\mathrm{SN}}\simeq2000\ \mathrm{yr}$,
we can neglect their energy losses. We thus solve Eqs.
(\ref{eq:dndt}) and (\ref{eq:dndt2}) without the energy loss term
and for a source function that provides protons of momentum
$\bar{p}_{\mathrm{o}}$ at the rate $\bar{Q}_{\mathrm{o}}$. The
proton energy distribution in the acceleration zone has then the
simple form
\begin{equation}\label{eq:nptp}
N_1(p,t)=\frac{\bar{Q}_{\mathrm{o}}\bar{t}_{\mathrm{a,o}}}{\bar{p}_{\mathrm{o}}}
\left(\frac{p}{\bar{p}_{\mathrm{o}}}\right)^{-s_1},
\end{equation}
in the momentum range $\bar{p}_{\mathrm{o}}<p<\bar{p}_{1}(t)$ and
$\bar{t}_{\mathrm{a,o}}\equiv
t_{\mathrm{acc}}(\bar{p}_{\mathrm{o}})$. The upper limit of the
proton distribution follows from the solution of Eq. (\ref{eq:char})
with $\alpha_1=0$ and is given by
\begin{equation}\label{eq:ptp}
\bar{p}_{1}(t)=\bar{p}_{\mathrm{o}}\left(1+\delta\frac{
t}{\bar{t}_{\mathrm{a,o}}}\right)^{\frac{1}{\delta}}.
\end{equation}
The distribution in the escape region is
\begin{equation}\label{eq:npt2p}
N_2(p,t)=\frac{3}{r-1}\frac{\bar{Q}_{\mathrm{o}}\,\left(t-\bar{t}_1(p)\right)}{\bar{p}_{\mathrm{o}}}\left(
\frac{p}{\bar{p}_{\mathrm{o}}}\right)^{-s_2},
\end{equation}
in the momentum range $\bar{p}_{\mathrm{o}}<p<\bar{p}_{1}(t)$, where
again $\bar{t}_1(p)$ is the inverse function of $\bar{p}_1(t)$. We
note that the power-law indices $s_1$, $s_2$ are the same as in the
leptonic case.

\section{Application to the supernova remnant \object{RX~J1713.7$-$3946}}\label{sec:apply}

We now apply the model developed in the previous section to the
supernova remnant \object{RX~J1713.7$-$3946}. We first consider the
case where only electrons are responsible for the observed emission,
that is we neglect the possible contribution from a hadronic
component. As discussed in section \ref{sec:model} there exist two
relativistic electron populations, one in the acceleration region
and one in the escape region behind the shock. These electrons will
produce synchrotron radiation in the magnetic field of the two
regions, assumed equal for simplicity. Electrons will also Compton
up-scatter the photons of the cosmic microwave and infrared
background radiation fields which are the same in both zones. For
the microwave photon field we take the standard values of
temperature $T_{\mathrm{MW}}=2.7\ \mathrm{K}$ and energy density
$U_{\mathrm{MW}}=0.25\ \mathrm{eV\ cm^{-3}}$ while for the infrared
field we assume $T_{\mathrm{IR}}=25\ \mathrm{K}$ and
$U_{\mathrm{IR}}=0.05\ \mathrm{eV\ cm^{-3}}$ \citep{aharonian05}. In
the possible emission processes we also include non-thermal
bremsstrahlung radiation and assume for the ambient hydrogen density
the average value $n\sim1\ \mathrm{cm}^{-3}$.

The photon spectrum derives by folding the two electron distribution
functions with the single particle emissivity for the synchrotron,
IC scattering and bremsstrahlung processes and thus depends on the
same parameters that determine the electron distribution function in
the acceleration and escape zones. These parameters are the
compression ratio of the shock $r$, the index of the diffusion
coefficient $\delta$, the magnetic field $B$, the maximum possible
electron energy $E_{\mathrm{e,max}}=p_{\mathrm{max}}c$ and the
normalization constant $Q_{\mathrm{o}}$ and can be determined from
the fit of the model to the observed spectrum. The injection
momentum of electrons does not alter the spectrum significantly and
so we take the arbitrary value $p_{\mathrm{o}}=5\
\mathrm{MeV}/\mathrm{c}$. We should also note that the calculation
of the photon spectrum is for the present day $t=t_{\mathrm{SN}}$.

\begin{figure}
\resizebox{\hsize}{!}{\includegraphics{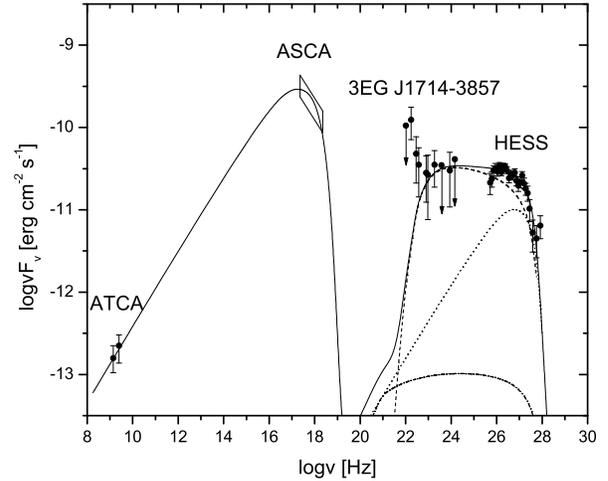}} \caption{The
multi-wavelength spectrum of \object{RX~J1713.7$-$3946} together
with the best-fit model with parameters given in text. We also show
the GeV upper limits from the nearby source
\object{3EG~J1714$-$3857}. The dashed line represents the emission
from neutral pions, the dotted line from IC scattering and the solid
line is the total emission. Also plotted is the bremsstrahlung
emission for the average ambient density $n=1\ \mathrm{cm^{-3}}$ as
the dash-dotted line.} \label{fit}
\end{figure}

The multi-wavelength spectrum of \object{RX~J1713.7$-$3946} is shown
in Fig.~\ref{fit}. Radio data from the northwest part of the remnant
taken with ATCA \citep{lazendic04} are shown here multiplied by a
factor of $2$ to cover the whole remnant. Likewise, the X-ray data
of ASCA \citep{enomoto02} are multiplied by a factor of $4$ so that
the total X-ray flux from the remnant is obtained \citep{tomida99}.
Finally, the TeV data are the recent observations of H.E.S.S. for
the entire remnant \citep{aharonian05}. We also plot the EGRET data
of the nearby source \object{3EG~J1714$-$3857}, since as emphasized
by \citet{reimer02} they set an upper limit to the GeV emission of
\object{RX~J1713.7$-$3946}.

We first examine a purely leptonic model in which the X-ray emission
is due to synchrotron radiation and the $\gamma$-ray emission is due
to IC scattering. From the observational data plotted in
Fig.~\ref{fit} we deduce that the X-ray emission peaks at
$\varepsilon_{\mathrm{max}}^{\mathrm{X}}\sim1\ \mathrm{keV}$ and the
$\gamma$-ray emission at
$\varepsilon_{\mathrm{max}}^{\mathrm{TeV}}\sim1\ \mathrm{TeV}$.
Since the peak frequencies scale as $E_{\mathrm{e,max}}^2 B$ and
$E_{\mathrm{e,max}}^2$ respectively (IC scattering takes place
mostly in the Thomson regime), we find that the magnetic field
should be $B\sim 40\ \mathrm{\mu G}$ and also that
$E_{\mathrm{e,max}}\sim45\ \mathrm{TeV}$. For these values, electron
cooling is important and the escaping particles have a broken
power-law distribution which gives the correct shape for the
extended $\gamma$-ray spectrum. However, the X-ray flux level is
then overestimated by a factor of 15 and one should use the
so-called magnetic filling factor to get the correct flux level.
This however cannot be done in our model since the emitted radiation
is calculated self-consistently from the particle losses. We thus
see that a synchrotron-IC model cannot explain at the same time both
the relative flux levels and the relative position of the peak
frequencies. The possibility that the $\gamma$-ray emission is due
to bremsstrahlung radiation can most probably be ruled out, since
the ambient hydrogen density required to fit the spectrum is
$n\sim300\ \mathrm{cm}^{-3}$, a value that might be too high. We
conclude that the observed radiation from this remnant cannot be of
purely leptonic origin.

We thus consider that protons are also accelerated in the remnant.
As a secondary electron population produced from charged pion decay
cannot explain the X-ray emission \citep{mastichiadis96}, primary
electrons are again required. However, in this case both electrons
and protons contribute to the $\gamma$-ray regime. The inclusion of
a relativistic proton population introduces one more free parameter,
the energy content in protons $\mathcal{E}_{\mathrm{p}}$ (related to
$\bar{Q}_{\mathrm{o}}$). For the minimum proton energy we adopt the
value $E_{\mathrm{p,min}}=2\ \mathrm{GeV}$ while the maximum proton
energy is determined from Eq. (\ref{eq:ptp}). For the emission from
neutral pions we use a $\delta$-function approximation.

We fit our model to the multi-wavelength spectrum of
\object{RX~J1713.7$-$3946} for the parameters $r$, $\delta$, $B$,
$E_{\mathrm{e,max}}$, $Q_{\mathrm{o}}$ and $\bar{Q}_{\mathrm{o}}$.
The resulting theoretical spectrum is shown in Fig.~\ref{fit}
together with the observational data. Although there is a small
disagreement with the first two H.E.S.S. points (at 215 GeV and 255
GeV), the fit is quite satisfactory and has a chi square
$\chi^{2}\simeq23$ with 22 degrees of freedom (when the last
H.E.S.S. point at 35 TeV is excluded from the fit). From the derived
values of $r=3.8$ and $\delta =0.3$ we can calculate the two
particle power-law indices $s_1=1.77$ and $s_2=2.07$. We note that
the value of $r$ implies a strong shock and the value of $\delta$ a
nearly Kolmogorov-type diffusion $K(p)\propto p^{0.3}$. The magnetic
field has the value $B=15\ \mathrm{\mu G}$. From this value and the
maximum electron energy $E_{\mathrm{e,max}}=100\ \mathrm{TeV}$ we
find that the cooling timescale is $t_{\mathrm{cool}}=500\
\mathrm{yr}$ and hence the energy losses in the escape region are
marginally important. We also deduce that electrons in the
acceleration zone are close to steady-state,
$p_1(t_{\mathrm{SN}})=0.7\ p_{\mathrm{max}}$, while those in the
escape zone have the broken power-law distribution with the break at
$p_{\mathrm{cool}}(t_{\mathrm{SN}})=20\ \mathrm{TeV}/\mathrm{c}=0.2\
p_{\mathrm{max}}$. For the protons the maximum energy at the present
day is $160\ \mathrm{TeV}$. The normalization constant
$Q_{\mathrm{o}}=6\ 10^{40}\ \mathrm{s}^{-1}$ determines the total
energy content in electrons from both zones
$\mathcal{E}_{\mathrm{e}}=3\ 10^{47}\ \mathrm{erg}$. The proton
normalization constant is $\bar{Q}_{\mathrm{o}}=2\ 10^{41}\
\mathrm{s}^{-1}$ and the corresponding energy content in protons is
$\mathcal{E}_{\mathrm{p}}=2\ 10^{50}\ \mathrm{erg}$, dominating the
total energy content in relativistic particles, i.e.
$\mathcal{E}_{\mathrm{tot}}\simeq\mathcal{E}_{\mathrm{p}}$. The
majority of this energy is carried by particles in the escape
region. For the typical value of supernova explosion energy of a few
times $10^{51}\ \mathrm{erg}$ we find that a fraction $\sim10\%$ of
it is needed in order to explain the non-thermal emission of
\object{RX~J1713.7$-$3946}.

From these parameters we find that the diffusion coefficient has the
form
\[K(p)=3\ 10^{26}\ \left(\frac{U_{\mathrm{sh}}}{6\ 10^3\ \mathrm{km}\ \mathrm{s}^{-1}}\right)^2\ \left(\frac{p}{100\ \mathrm{TeV}\ \mathrm{c}^{-1}}\right)^{0.3}\ \mathrm{cm}^2\ \mathrm{s}^{-1}\]
and is greater than the lower limit of Bohm diffusion for all
particle momenta less than
\[p<170\ \left(\frac{U_{\mathrm{sh}}}{6\ 10^3\ \mathrm{km}\ \mathrm{s}^{-1}}\right)^{20/7}\ \mathrm{TeV}\ \mathrm{c}^{-1}.\]
If we thus adopt for the shock velocity the value
$U_{\mathrm{sh}}=6\ 10^3\ \mathrm{km}\ \mathrm{s}^{-1}$ then the
diffusion coefficient we derive does not violate the lower limit of
Bomh diffusion for any value of particle momentum in our model.

\section{Summary and Discussion}\label{sec:discuss}

In the present paper we have attempted a fit to the multi-wavelength
spectrum of the supernova remnant \object{RX~J1713.7$-$3946} based
on the box model of \citet{drury99}. According to this, particles
diffuse in an energy-dependent region around the shock front of the
remnant and are accelerated by the first order Fermi mechanism. The
particle distribution functions (both for electrons and protons)
come as solutions to two coupled partial differential equations, one
characterizing the particles inside the acceleration region and the
other the ones that have escaped from it. Consequently, one can
calculate the radiated photon spectra by folding the derived
particle distribution functions with the single particle emissivity
for the radiative processes assumed. For the case of electrons the
corresponding radiative processes are synchrotron, IC scattering and
bremsstrahlung emission, while protons produce radiation only
through the decay of pions.

We found that, within the strict assumptions of the box model, a
purely leptonic origin of the radiation cannot explain the
observations. The leptonic model is viable only in the case where
the magnetic field in the escape region has a value large enough as
to allow for substantial particle cooling during the lifetime of the
remnant. The escaping particles then have a broken power-law
distribution that can explain the extended $\gamma$-ray spectrum of
\object{RX~J1713.7$-$3946}. For such a high value of the magnetic
field though, the level of X-ray emission exceeds significantly the
observed one. We thus reach to the same conclusion with all the
other leptonic models used for this source, i.e. that we cannot
explain at the same time the relative position of the synchrotron
and IC peaks and their relative fluxes, unless we use the filling
factor. We should emphasize however that the above conclusions are
valid under the assumptions of a spatially averaged model and a
plane-parallel shock geometry. In a model where the spatial
dependence is treated in detail one could include other effects like
variable injection rate at different supernova radii and/or
expansion losses, although the error from the latter should not be
important. Whether a purely leptonic model would be more successful
in such a case is something that remains to be investigated.

The inclusion of the relativistic hadronic population provides an
additional component in $\gamma$-rays and a satisfactory fit can
then be obtained. However, the hadronic component can be treated
only approximately in our model since the test particle
approximation that we used is not strictly valid and the shock gets
modified by the hadrons. To take this into account one should use
calculations based on numerical models.

From the fit to the spectrum of \object{RX~J1713.7$-$3946} we also
deduced the dependence of the diffusion coefficient on energy. We
found that a nearly Kolmogorov-type diffusion coefficient
($\delta=0.3$) can reproduce the observed spectrum quite well. The
1$\sigma$ estimate for the parameter $\delta$ is well concentrated
around this value, $0.1\lesssim \delta\lesssim 0.5$. This is because
a change in $\delta$ mostly influences the maximum electron and
proton energies and these are well defined from the observed
spectrum. We note that this range of $\delta$ allows marginally for
a Kraichnan-type turbulence spectrum, while the case of Bohm
diffusion is not favored by our results.

We conclude that the model we used, despite certain assumptions, has
two main advantages over other models, in that a) the particle
distribution functions are calculated from a theoretical model and
are not assumed a priori and b) it can give a method of estimating
the energy dependence of the diffusion coefficient in supernova
remnants.

\begin{acknowledgements}
The authors would like to thank the anonymous referee for valuable
comments as well as F. Aharonian for useful discussions and N.
Vlahakis for making comments that helped improve the manuscript. The
project is co-funded by the European Social Fund and National
Resources -- (EPEAEK II) PYTHAGORAS.
\end{acknowledgements}

\bibliography{References}

\end{document}